\documentclass[twocolumn,pre]{revtex4}

\usepackage{dcolumn}
\usepackage{amsmath}

\usepackage{graphicx}
\usepackage{calrsfs}
\usepackage{bm}

\begin{document}

\renewcommand{\d}{\mathrm{d}}
\newcommand{\Ord}{\mathrm{O}}
\newcommand{\e}{\mathrm{e}}
\newcommand{\ii}{\mathrm{i}}
\newcommand{\half}{\mbox{$\frac12$}}
\newcommand{\set}[1]{\lbrace#1\rbrace}
\newcommand{\av}[1]{\left\langle#1\right\rangle}
\newcommand{\eref}[1]{(\ref{#1})}
\newcommand{\etal}{{\it{}et~al.}}
\newcommand{\defn}{\textit}
\newcommand{\cD}{\mathcal{D}}
\newcommand{\cG}{\mathcal{G}}
\newcommand{\cH}{\mathcal{H}}
\newcommand{\vphi}{\bm{\phi}}
\newcommand{\Tr}{\mathop{\rm Tr}}
\newcommand{\thetain}{\theta^\mathrm{in}}
\newcommand{\thetaout}{\theta^\mathrm{out}}
\newcommand{\kin}{k^\mathrm{in}}
\newcommand{\kout}{k^\mathrm{out}}
\newcommand{\hlparam}{\alpha}
\newcommand{\tsparam}{\alpha}

\newlength{\figurewidth}
\setlength{\figurewidth}{0.95\columnwidth}
\setlength{\parskip}{0pt}
\setlength{\tabcolsep}{6pt}
\setlength{\arraycolsep}{2pt}

\title{The statistical mechanics of networks}
\author{Juyong Park}
\affiliation{Department of Physics and Center for the Study of Complex
Systems,\\
University of Michigan, Ann Arbor, MI 48109--1120}
\author{M. E. J. Newman}
\affiliation{Department of Physics and Center for the Study of Complex
Systems,\\
University of Michigan, Ann Arbor, MI 48109--1120}

\begin{abstract}
We study the family of network models derived by requiring the expected
properties of a graph ensemble to match a given set of measurements of a
real-world network, while maximizing the entropy of the ensemble.  Models
of this type play the same role in the study of networks as is played by
the Boltzmann distribution in classical statistical mechanics; they offer
the best prediction of network properties subject to the constraints
imposed by a given set of observations.  We give exact solutions of models
within this class that incorporate arbitrary degree distributions and
arbitrary but independent edge probabilities.  We also discuss some more
complex examples with correlated edges that can be solved approximately or
exactly by adapting various familiar methods, including mean-field theory,
perturbation theory, and saddle-point expansions.
\end{abstract}
\pacs{89.75.Hc, 87.23.Ge, 89.20.Hh, 05.10.-a}
\maketitle

\section{Introduction}
The last few years have seen the publication of a large volume of work in
the physics literature on networks of various kinds, particularly computer
and information networks like the Internet and world wide web, biological
networks such as food webs and metabolic networks, and social
networks~\cite{Strogatz01,AB02,DM02,Newman03d}.  This work has been divided
between empirical studies of the structure of particular networks and
theoretical studies focused largely on the creation of mathematical and
computational models.  The construction of network models is the topic of
this paper.

Models of networks can help us to understand the important features of
network structure and the interplay of structure with processes that take
place on networks, such as the flow of traffic on the Internet or the
spread of a disease over a social network.  Most network models studied in
the physics community are of a practical sort.  Typically one wishes to
create a network that displays some feature or features observed in
empirical studies.  The principal approach is to list possible mechanisms
that might be responsible for creating those features and then make a model
incorporating some or all of those mechanisms.  One then either examines
the networks produced by the model for rewarding similarity to the
real-world systems they are supposed to mimic, or uses them as a substrate
for further modeling, for example of traffic flow or disease spread.
Classic examples of models of this kind are the small-world
model~\cite{WS98} and the many different preferential attachment
models~\cite{BA99b,KRL00,DMS00}, which model network transitivity and
power-law degree distributions respectively.

However, there is another possible approach to the modeling of networks,
which has been pursued comparatively little so far.  An instructive analogy
can be made here with theories of gases.  There are (at least) two
different general theories of the properties of gases.  Kinetic theory
explicitly models collections of individual atoms, their motions and
collisions, and attempts to calculate overall properties of the resulting
system from basic mechanical principles.  Pressure, for instance, is
calculated from the mean momentum transfered to the walls of a container by
bombarding atoms.  Kinetic theory is well motivated, easy to understand,
and makes good sense to physicists and laymen alike.  However, kinetic
theory rapidly becomes complex and difficult to use if we attempt to make
it realistic by the inclusion of accurate intermolecular potentials and
similar features.  In practice, kinetic theory models either make only
rather rough and uncontrolled predictions, or they rely on large-scale
computer simulation to achieve accuracy.

If one wants a good calculational tool for studying the properties of
gases, therefore, one does not use kinetic theory.  Instead, one uses
statistical mechanics.  Although certainly less intuitive, statistical
mechanics is based on rigorous probabilistic arguments and gives accurate
and reliable answers for an enormous range of problems, including many,
such as problems concerning solids, for which kinetic theory is
inapplicable.  Equilibrium statistical mechanics provides a general
framework for reasoning and a powerful calculational tool for very many
problems in statistical physics.

Here we argue that the current commonly used models of networks are akin to
kinetic theory.  They posit plausible mechanisms or dynamics, and produce
results in qualitative agreement with reality, at least in some respects.
They are easy to understand and give us good physical insight.  However,
like kinetic theory, they do not make quantitatively accurate predictions
and provide no overall framework for modeling, each network model instead
concentrating on explaining one or a few features of the system of interest.

In this paper we discuss exponential random graphs, which are to networks
as statistical mechanics is to the study of gases---a well-founded general
theory with true predictive power.  These advantages come at a price:
exponential random graphs are both mathematically and conceptually
sophisticated, and their understanding demands some effort of the reader.
We believe this effort to be more than worthwhile, however.  Theoretical
techniques based on solid statistical foundations and capable of
quantitative predictions have been of extraordinary value in the study of
fluid, solid state, and other physical systems, and there is no reason to
think they will be any less valuable for networks.

We are by no means the first authors to study exponential random graphs,
although our approach is different from that taken by others.  Exponential
random graphs were first proposed in the early 1980s by Holland and
Leinhardt~\cite{HL81}, building on statistical foundations laid by
Besag~\cite{Besag74}.  Substantial further developments were made by Frank
and Strauss~\cite{Frank81,FS86,Strauss86}, and continued to be made by
others throughout the 1990s~\cite{WP96,AWC99}.  In recent years a number of
physicists, including ourselves, have made theoretical studies of specific
cases~\cite{BL02,PN03,PN04,BJK04a,BJK04b,PDFV04}.  Today, exponential
random graphs are in common use within the statistics and social network
analysis communities as a practical tool for modeling networks and several
standard computer tools are available for simulating and manipulating them,
including Prepstar, ERGM, and Siena~\cite{Snijders02}.

In this paper we aim to do a number of things.  First, we place exponential
random graph models on a firm physical foundation, showing that they can be
derived from first principles using maximum entropy arguments.  In doing
so, we argue that these models are not merely an \textit{ad hoc}
formulation studied primarily for their mathematical convenience, but a
true and correct extension of the statistical mechanics of Boltzmann and
Gibbs to the network world.

Second, we take an almost entirely analytic approach in our work, by
contrast with the numerical simulations that form the core of most previous
studies.  We show that the analytic techniques of equilibrium statistical
mechanics are ideally suited to the study of these models and can shed much
light on their structure and behavior.  Throughout the paper we give
numerous examples of specific models that are solvable either exactly or
approximately, including several that have a long history in network
analysis.  Nonetheless, the particular examples studied in this paper form
only a tiny fraction of the possibilities offered by this class of models.
There are many intriguing avenues for future research on exponential random
graphs that are open for exploration, and we highlight a number of these
throughout the paper.

\section{Exponential random graphs}
The typical scenario addressed in the creation of a network model is this:
one has measurements of a number of network properties for a real-world
network or networks, such as number of vertices or edges, vertex degrees,
clustering coefficients, correlation functions, and so forth, and one
wishes to make a model network that has the same or similar values of these
properties.  For instance, one might find that a network has a degree
sequence with a power-law distribution and wish to create a model network
that shows the same power law.  Or one might measure a high clustering
coefficient in a network and wish to build a model network with similarly
high clustering.

Essentially all models considered in modern work, and indeed as far back as
the 1950s and 1960s, have been ensemble models, meaning that a model is
defined to be not a single network, but a probability distribution over
many possible networks.  We adopt this approach here as well.  Our goal
will be to choose a probability distribution such that networks that are a
better fit to observed characteristics are accorded higher probability in
the model.

Consider a set $\cG$ of graphs.  One can use any set~$\cG$, but in most of
the work described in this paper $\cG$ will be the set of all simple graphs
without self-loops on $n$ vertices.  (A simple graph is a graph having at
most a single edge between any pair of vertices.  A self-loop is an edge
that connects a vertex to itself.)  Certainly there are many other possible
choices and we consider some of the others briefly in
Sections~\ref{fermibose} and~\ref{fixed}.  The graphs can also be either
directed or undirected and we consider both in this paper, although most of
our time will be spent on the undirected case.

Suppose we have a collection of graph observables $\set{x_i}$, $i=1\ldots
r$, that we have measured in empirical observation of some real-world
network or networks of interest to us.  We will, for the sake of
generality, assume that we have an estimate $\av{x_i}$ of the expectation
value of each observable.  In practice it is often the case that we have
only one measurement of an observable.  For instance, we have only one
Internet, and hence only one measurement of the clustering coefficient of
the Internet.  In that case, however, our best estimate of the expectation
value of the clustering coefficient is simply equal to the one measurement
that we have.

Let $G\in\cG$ be a graph in our set of graphs and let $P(G)$ be the
probability of that graph within our ensemble.  We would like to choose
$P(G)$ so that the expectation value of each of our graph observables
$\set{x_i}$ within that distribution is equal to its observed value, but
this is a vastly underdetermined problem in most cases; the number of
degrees of freedom in the definition of the probability distribution is
huge compared to the number of constraints imposed by our observations.
Problems of this type however are commonplace in statistical physics and we
know well how to deal with them.  The best choice of probability
distribution, in a sense that we will make precise in a moment, is the one
that maximizes the Gibbs entropy
\begin{equation}
S = - \sum_{G\in\cG} P(G) \ln P(G),
\end{equation}
subject to the constraints
\begin{equation}
\sum_G P(G) x_i(G) = \av{x_i},
\label{constraints}
\end{equation}
plus the normalization condition
\begin{equation}
\sum_G P(G) = 1.
\end{equation}
Here $x_i(G)$ is the value of $x_i$ in graph~$G$.

Introducing Lagrange multipliers $\alpha, \set{\theta_i}$, we then find
that the maximum entropy is achieved for the distribution satisfying
\begin{eqnarray}
\hspace{-2em} {\partial\over\partial P(G)} \biggl[ S
  &+& \alpha \biggl( 1 - \sum_G P(G) \biggr) \nonumber\\
  &+& \sum_i \theta_i
      \biggl( \av{x_i} - \sum_G P(G) x_i(G) \biggr) \biggr] = 0
\end{eqnarray}
for all graphs~$G$.  This gives
\begin{equation}
\ln P(G) + 1 + \alpha + \sum_i \theta_i x_i(G) = 0,
\end{equation}
or equivalently
\begin{equation}
P(G) = {\e^{-H(G)}\over Z},
\label{defspg}
\end{equation}
where $H(G)$ is the graph Hamiltonian
\begin{equation}
H(G) = \sum_i \theta_i x_i(G)
\label{defsh}
\end{equation}
and $Z$ is the partition function
\begin{equation}
Z = \e^{\alpha+1} = \sum_G \e^{-H(G)}.
\label{defsz}
\end{equation}
Equations~\eref{defspg} to~\eref{defsz} define the exponential random graph
model.  The exponential random graph is the distribution over a specified
set of graphs that maximizes the entropy subject to the known constraints.
It is also the exact analogue for graphs of the Boltzmann distribution of a
physical system over its microstates at finite temperature.

Using the exponential random graph model involves performing averages over
the probability distribution~\eref{defspg}.  The expected value of any
graph property~$x$ within the model is simply
\begin{equation}
\av{x} = \sum_G P(G) x(G).
\label{avx}
\end{equation}
The exponential random graph, like all such maximum entropy ensembles,
gives the best prediction of an unknown quantity $x$, given a set of known
quantities, Eq.~\eref{constraints}.  In this precise sense, the exponential
random graph is the best ensemble model we can construct for a network
given a particular set of observations.

In many cases we may not need to perform the sum~\eref{avx}; often we need
only perform the partition function sum, Eq.~\eref{defsz}, and the values
of other sums can then be deduced by taking appropriate derivatives.  Just
as in conventional equilibrium statistical mechanics, however, performing
even the partition function sum analytically may not be easy.  Indeed in
some cases it may not be possible at all, in which case one may have to
turn to Monte Carlo simulation, to which the model lends itself admirably.
As we show in this paper however, there are a variety of tools one can
employ to get exact or approximate analytic solutions in cases of interest,
including mean-field theory, algebraic transformations, and diagrammatic
perturbation theory.

\section{Simple examples}
Before delving into the more complicated calculations, let us illustrate
the use of exponential random graphs with some simple examples.

\subsection{Random graphs}
Consider first what is perhaps the simplest of exponential random graphs,
at least for the case of fixed number of vertices~$n$ considered here.

Suppose we know only the expect number of edges $\av{m}$ that our network
should have.  In that case the Hamiltonian takes the simple form
\begin{equation}
H(G) = \theta m(G).
\label{randomgraph}
\end{equation}
We can think of the parameter $\theta$ as either a field coupling to the
number of edges, or alternatively as an inverse temperature.

Let us evaluate the partition function for this Hamiltonian for the case of
an ensemble of simple undirected graphs on $n$ vertices without self-loops.
We define the \defn{adjacency matrix} $\bm{\sigma}$ to be the symmetric
$n\times n$ matrix with elements
\begin{equation}
\sigma_{ij} = \biggl\lbrace\begin{array}{ll}
                 1 & \qquad\mbox{if $i$ is connected to $j$,} \\
                 0 & \qquad\mbox{otherwise}.
              \end{array}
\label{adjacency}
\end{equation}
Then the number of edges is $m=\sum_{i<j} \sigma_{ij}$, and the partition
function is
\begin{eqnarray}
Z &=& \sum_G \e^{-H}
   =  \sum_{\set{\sigma_{ij}}}
      \exp\biggl(-\theta \sum_{i<j} \sigma_{ij} \biggr) \nonumber\\
  &=& \prod_{i<j}\hspace{4pt}\sum_{\sigma_{ij}=0}^1 \e^{-\theta\sigma_{ij}}
   =  \prod_{i<j} \bigl( 1 + \e^{-\theta} \bigr) \nonumber\\
  &=& \bigl[ 1 + \e^{-\theta} \bigr]^{n\choose2}.
\label{bernoulli}
\end{eqnarray}

It is convenient to define the free energy
\begin{equation}
F = -\ln Z,
\label{defsf}
\end{equation}
which in this case is
\begin{equation}
F = -{n\choose 2} \ln \bigl( 1 + \e^{-\theta} \bigr).
\label{frg}
\end{equation}
(Note that the free energy is extensive not in the number of vertices~$n$,
but in the number ${n\choose2}$ of pairs of vertices, since this is the
number of degrees of freedom in the model.)  Then, for instance, the
expected number of edges in the model is
\begin{eqnarray}
\av{m} &=& {1\over Z} \sum_G m \e^{-H}
        =  -{1\over Z} {\partial Z\over\partial\theta}
        =  {\partial F\over\partial\theta} \nonumber\\
       &=& {n\choose2} {1\over\e^\theta + 1}.
\end{eqnarray}
Conventionally we re-express the parameter~$\theta$ in terms of
\begin{equation}
p = {1\over\e^\theta + 1},
\label{defsp}
\end{equation}
so that $\av{m}={n\choose2}p$.

The probability $P(G)$ of a graph in this ensemble can be written
\begin{equation}
P(G) = {\e^{-H}\over Z}
     = {\e^{-\theta m}\over \bigl[ 1 + \e^{-\theta} \bigr]^{n\choose2}}
     = p^m (1-p)^{{n\choose2}-m}.
\end{equation}
In other words, $P(G)$ is simply the probability for a graph in which each
of the ${n\choose2}$ possible edges appears with independent
probability~$p$.

This model is known as the Bernoulli random graph, or often just the random
graph, and was introduced, in a completely different fashion, by Solomonoff
and Rapoport~\cite{SR51} in 1951 and later famously studied by Erd\H{o}s
and R\'enyi~\cite{ER59,ER60}.  Today it is one of the best studied of graph
models, although, as many authors have pointed out, it is not a good model
of most real-world networks~\cite{Strogatz01,DM02,WS98}.  One way in which
its inadequacy shows, and one that has been emphasized heavily in networks
research in the last few years, is its degree distribution.  Since each
edge in the model appears with independent probability~$p$, the degree of a
vertex, i.e.,~the number of edges attached to that vertex, follows a
binomial distribution, or a Poisson distribution in the limit of large~$n$.
Most real-world networks however have degree distributions that are far
from Poissonian, typically being highly right-skewed, with a small
proportion of vertices having very high degree.  Some of the most
interesting networks, including the Internet and the world wide web, appear
to have degree distributions that follow a power
law~\cite{BA99b,FFF99,Kleinberg99b}.  In the next section we discuss what
happens when we incorporate observations like these into our models.

\subsection{Generalized random graphs}
\label{grg}
Suppose then that rather than just measuring the total number of edges in a
network, we measure the degrees of all the vertices.  Let us denote by
$k_i$ the degree of vertex~$i$.  The complete set~$\set{k_i}$ is called the
\defn{degree sequence} of the network.  Note that we do not need to specify
independently the number of edges~$m$ in the network, since $m=\half\sum_i
k_i$ for an undirected graph.

The exponential random graph model appropriate to this set of observations
is the model having Hamiltonian
\begin{equation}
H = \sum_i \theta_i k_i,
\label{genrg}
\end{equation}
where we now have one parameter $\theta_i$ for each vertex~$i$.  Noting
that $k_i=\sum_j\sigma_{ij}$, this can also be written
\begin{equation}
H = \sum_{ij} \theta_i \sigma_{ij}
  = \sum_{i<j} (\theta_i+\theta_j) \sigma_{ij}.
\end{equation}
Then the partition function is
\begin{eqnarray}
Z &=& \!\!\sum_{\set{\sigma_{ij}}}
      \exp\biggl(-\sum_{i<j} (\theta_i+\theta_j)\sigma_{ij} \biggr)
   =  \prod_{i<j}\hspace{4pt}\sum_{\sigma_{ij}=0}^1
      \e^{-(\theta_i+\theta_j)\sigma_{ij}} \nonumber\\
  &=& \prod_{i<j} \bigl( 1 + \e^{-(\theta_i+\theta_j)} \bigr),
\label{genrgpart}
\end{eqnarray}
and the free energy is
\begin{equation}
F = -\sum_{i<j} \ln \bigl( 1 + \e^{-(\theta_i+\theta_j)} \bigr).
\label{fgenrg}
\end{equation}

More generally we could specify a Hamiltonian
\begin{equation}
H = \sum_{i<j} \Theta_{ij} \sigma_{ij},
\label{gengenrg}
\end{equation}
with a separate parameter $\Theta_{ij}$ coupling to each
edge~\cite{Frank81}.  Then
\begin{equation}
Z = \prod_{i<j} \bigl( 1 + \e^{-\Theta_{ij}} \bigr),\quad
F = -\sum_{i<j} \ln \bigl( 1 + \e^{-\Theta_{ij}} \bigr).
\label{gengenrgpart}
\end{equation}
This allows us for example to calculate the probability of occurrence
$p_{ij}$ of an edge between vertices $i$ and~$j$:
\begin{equation}
p_{ij} = \av{\sigma_{ij}} = {\partial F\over\partial\Theta_{ij}}
       = {1\over\e^{\Theta_{ij}} + 1}.
\label{edgeprob1}
\end{equation}
The model of Eq.~\eref{genrg} is the special case in which
$\Theta_{ij}=\theta_i+\theta_j$ and the normal (Bernoulli) random graph of
Eq.~\eref{bernoulli} corresponds to the case in which the parameters
$\Theta_{ij}$ are all equal.

Sometimes it is convenient to specify not a degree sequence but a
probability distribution over vertex degrees.  This can be achieved by
specifying an equivalent distribution over the parameters $\theta_i$
in~\eref{genrg}.  Let us define $\rho(\theta)\>\d\theta$ to be the
probability that the parameter $\theta$ for a vertex lies in the range
$\theta$ to $\theta+\d\theta$.  Then, averaging over the disorder so
introduced, the free energy, Eq.~\eref{fgenrg}, becomes
\begin{eqnarray}
F &=& - \int \rho(\theta_1)\,\d\theta_1\ldots\rho(\theta_n)\,\d\theta_n
        \sum_{i<j} \ln \bigl( 1 + \e^{-(\theta_i+\theta_j)} \bigr)
        \nonumber\\
  &=& - {n\choose2} \iint \ln \bigl( 1 + \e^{-(\theta+\theta')} \bigr)
        \rho(\theta)\rho(\theta')\>\d\theta\>\d\theta'.
\label{distf}
\end{eqnarray}
The part of this free energy due to a single vertex with field
parameter~$\theta$ is
\begin{equation}
{1\over n}\,{\delta F\over\delta\rho(\theta)} =
  - (n-1) \int \ln \bigl( 1 + \e^{-(\theta+\theta')} \bigr)
    \rho(\theta')\>\d\theta',
\end{equation}
and the expected degree of vertex~$i$ with field~$\theta_i$ is the
derivative of this with respect to~$\theta$, evaluated at~$\theta_i$:
\begin{eqnarray}
\av{k_i} &=& - (n-1) \biggl[ {\partial\over\partial\theta}\!
             \int \ln \bigl( 1 + \e^{-(\theta+\theta')} \bigr)\,
             \rho(\theta')\,\d\theta' \biggr]_{\theta=\theta_i} \nonumber\\
         &=& (n-1) \int {\rho(\theta')\,\d\theta'\over
                         \e^{\theta_i+\theta'}+1}.
\label{distkav}
\end{eqnarray}
By a judicious choice of $\rho(\theta)$ we can then produce the desired
degree distribution.  (See also Sec.~\ref{classical}.)  We studied this
model in a previous paper~\cite{PN03}, as a model for degree correlations
in the Internet and other networks.

We could alternatively specify a probability distribution $\rho(\Theta)$
for the parameters $\Theta_{ij}$ in~\eref{gengenrg} that couple to
individual edges.  Or, taking the developments a step further, one could
define joint distributions for the $\Theta_{ij}$ on different edges,
thereby introducing correlations of quite general kinds between the edges
in the model.  There are enormous possibilities to be explored in this
regard, but we pass over them for now, our interests in the present paper
lying in other directions.

One can calculate many other properties of our models.  For example, for
the model of Eq.~\eref{genrg}, one can calculate the expectation value of
any product of vertex degrees from an appropriate derivative of the
partition function:
\begin{equation}
\av{k_i k_j \ldots} = {1\over Z} \biggl[ {\partial\over\partial\theta_i}
                      {\partial\over\partial\theta_j} \ldots \biggr] Z.
\label{disconnected}
\end{equation}
Such derivatives are correlation functions of degrees within the model.
Similarly, derivatives of the free energy give the connected correlation
functions:
\begin{subequations}
\begin{eqnarray}
{\partial F\over\partial\theta_i} &=& \av{k_i},\\
{\partial^2 F\over\partial\theta_i\partial\theta_j}
  &=& \av{k_ik_j}_c = \av{k_ik_j} - \av{k_i}\av{k_j},\\
{\partial^3 F\over\partial\theta_i\partial\theta_j\partial\theta_l}
  &=& \av{k_ik_jk_l}_c \nonumber\\
  &=& \av{k_ik_jk_l} - \av{k_ik_j}_c\av{k_l} - \av{k_jk_l}_c\av{k_i}
      \nonumber\\
  & & {} - \av{k_lk_i}_c\av{k_j} - \av{k_i}\av{k_j}\av{k_l},
\end{eqnarray}
\label{concorrelation}
\end{subequations}
and so forth.

For instance, the two-vertex connected correlation is
\begin{equation}
\av{k_ik_j}_c = \left\lbrace\begin{array}{ll}
  {\e^{\theta_i+\theta_j}\over(\e^{\theta_i+\theta_j}+1)^2} &
  \qquad\mbox{for $i\ne j$,} \\
  (n-1) {\e^{2\theta_i}\over(\e^{2\theta_i}+1)^2} &
  \qquad\mbox{for $i=j$.}
\end{array}\right.
\end{equation}
For the case of the Bernoulli random graph, which has all $\theta_i$ equal,
this gives $\av{k_ik_j}_c = p(1-p)$ for $i\ne j$, where we have made use of
Eq.~\eref{defsp}.  Thus the degrees of vertices in the random graph are in
general positively correlated.  One can understand this as an effect of the
one edge that potentially connects the two vertices $i$ and~$j$.  The
presence or absence of this edge introduces a correlation between the two
degrees.  (For a sparse graph, in which $p=\Ord(n^{-1})$, the correlation
disappears in the limit of large graph size.)

In order to measure some quantities within exponential random graph models,
it may be necessary to introduce additional terms into the Hamiltonian.
For instance, to find the expectation value of the clustering
coefficient~$C$~\cite{WS98}, one would like to evaluate
\begin{equation}
\av{C} = {\sum_G C(G) \e^{-H}\over Z},
\end{equation}
which we can do by introducing an extra term linear in the clustering
coefficient in the Hamiltonian.  To measure clustering in the network of
Eq.~\eref{genrg}, for example, we could define
\begin{equation}
H = \sum_i \theta_i k_i + \gamma C.
\end{equation}
Then
\begin{equation}
\av{C} = {\partial F\over\partial\gamma}\biggr|_{\gamma=0}.
\end{equation}
Thus it is important, even in the simplest of cases, to be able to solve
more general models, and much of the rest of the paper is devoted to the
development of techniques to do this.

\subsection{Directed graphs}
\label{directedgraphs}
Before we look at more complicated Hamiltonians, let us look briefly at
what happens if we change the graph set~$\cG$ over which our sums are
performed.  The first case we examine is that of directed graphs.  We
define $\cG$ to be the set of all simple loopless directed graphs, which is
parameterized by the asymmetric adjacency matrix
\begin{equation}
\sigma_{ij} = \biggl\lbrace\begin{array}{ll}
                 1 & \qquad\mbox{if there is an edge from $j$ to~$i$,} \\
                 0 & \qquad\mbox{otherwise}.
              \end{array}
\label{adjacencydirected}
\end{equation}
Thus, for instance, the Hamiltonian $H=\theta m$ gives rise to a partition
function
\begin{equation}
Z = \prod_{i\ne j}\hspace{4pt}\sum_{\sigma_{ij}=0}^1 \e^{-\theta\sigma_{ij}}
  = \bigl[ 1 + \e^{-\theta} \bigr]^{2{n\choose2}}
\label{directedz}
\end{equation}
and a corresponding free energy.

The directed equivalent of the more general model of Eq.~\eref{genrg} in
which we can control the degree of each vertex is a model that now has two
separate parameters for each vertex, $\thetain_i$ and $\thetaout_i$, that
couple to the in- and out-degrees:
\begin{equation}
H = \sum_i \bigl(\thetain_i\kin_i + \thetaout_i\kout_i \bigr).
\end{equation}
Then the partition function and free energy are
\begin{eqnarray}
Z &=& \prod_{i\ne j} \bigl(1+\e^{-(\thetain_i+\thetaout_j)} \bigr) \\
F &=& -\sum_{i\ne j} \ln \bigl(1+\e^{-(\thetain_i+\thetaout_j)} \bigr).
\end{eqnarray}

From these we can calculate the expected in- and out-degree of a vertex:
\begin{eqnarray}
\av{\kin_i}  &=& {\partial F\over\partial\thetain_i}
              =  \sum_{j(\ne i)} {1\over\e^{(\thetain_i+\thetaout_j)}+1},\\
\av{\kout_i} &=& {\partial F\over\partial\thetaout_i}
              =  \sum_{j(\ne i)} {1\over\e^{(\thetain_j+\thetaout_i)}+1}.
\end{eqnarray}
We note that $\sum_i\av{\kin_i} = \sum_i\av{\kout_i}$, as must be the case
for all directed graphs, since every edge on such a graph must both start
and end at exactly one vertex.

We can also define a probability distribution $\rho(\thetain,\thetaout)$
for the fields on the vertices, and the developments generalize
Eqs.~(\ref{distf}--\ref{distkav}) in a natural fashion.

We give a more complex example of a directed graph model in
Section~\ref{reciprocity}, where we derive a solution to the reciprocity
model of Holland and Leinhardt~\cite{HL81} using perturbative methods.

\subsection{Fermionic and bosonic graphs}
\label{fermibose}
It will by now have occurred to many readers that results like
Eqs.~\eref{fgenrg} and~\eref{distkav} bear a similarity to corresponding
results from traditional statistical mechanics for systems of
non-interacting fermions.  We can look upon the edges in our networks as
being like particles in a quantum gas and pairs of vertices as being like
single-particle states.  Simple graphs then correspond to the case in which
each single-particle state can be occupied by at most one particle, so it
should come as no surprise that the results look similar to a system
obeying the Pauli exclusion principle.

Not all networks need have only a single edge between any pair of vertices.
Some can have multiple edges or \defn{multiedges}.  The world wide web is
an example---there can be and frequently is more than one link from one
page to another.  The Internet, airline networks, metabolic networks,
neural networks, citation networks, and collaboration networks are other
examples of networks that can exhibit multiedges.  There is no problem
generalizing our exponential random graphs to this case and, as we might
expect, it gives rise to a formalism that resembles the theory of bosons.

Let us define our set of graphs~$\cG$ to be the set of all undirected
graphs with any number of edges between any pair of vertices (but still no
self-edges, although there is no reason in principle why these cannot be
included as well).  Taking for example the Hamiltonian,
Eq.~\eref{gengenrg}, and generalizing the adjacency matrix,
Eq.~\eref{adjacency}, so that $\sigma_{ij}$ is now equal to the
\emph{number} of edges between $i$ and $j$, we have
\begin{eqnarray}
Z &=& \sum_{\set{\sigma_{ij}}}
      \exp \biggl( -\sum_{i<j} \Theta_{ij}\sigma_{ij} \biggr)
   =  \prod_{i<j}\hspace{4pt}
      \sum_{\sigma_{ij}=0}^\infty \e^{-\Theta_{ij}\sigma_{ij}}
      \nonumber\\
  &=& \prod_{i<j} {1\over1 - \e^{-\Theta_{ij}}},
\end{eqnarray}
and
\begin{equation}
F = \sum_{i<j} \ln\bigl( 1 - \e^{-\Theta_{ij}} \bigr).
\label{bosonic}
\end{equation}
The equivalent of the probability $p_{ij}$ of an edge appearing in the
fermionic case is now the expected number of edges $n_{ij}$ between
vertices $i$ and~$j$, which is given by
\begin{equation}
n_{ij} = \av{\sigma_{ij}} = {\partial F\over \partial \Theta_{ij}} =
{1\over\e^{\Theta_{ij}} - 1}.
\end{equation}
Note that this quantity diverges if we allow $\Theta_{ij}\to0$, a
phenomenon related to Bose-Einstein condensation in ordinary Bose gases.

For the special cases of Eqs.~\eref{genrg} and~\eref{randomgraph}, we have
\begin{equation}
F = \sum_{i<j} \ln \bigl( 1 - \e^{-(\theta_i+\theta_j)} \bigr),\qquad
n_{ij} = {1\over\e^{\theta_i+\theta_j}-1}
\end{equation}
and
\begin{equation}
F = {n\choose2} \ln \bigl( 1 - \e^{-\theta} \bigr),\qquad
n_{ij} = {1\over\e^\theta-1},
\end{equation}
respectively.  The connected correlation between the degrees of any two
vertices in the latter case is
\begin{equation}
\av{k_ik_j}_c = {\partial^2 F\over\partial\theta^2}
              = {\e^\theta\over(\e^\theta-1)^2},
\end{equation}
for $i\ne j$.  Thus the degrees are again positively correlated and the
correlation diverges as $\theta\to0$.

\subsection{The sparse or classical limit}
\label{classical}
In most real-world networks the number of edges~$m$ is quite small.
Typically $m$ is of the same order as~$n$, rather than being of
order~$n^2$.  Such graphs are said to be \defn{sparse}.  (One possible
exception is food webs, which appear to be dense, having
$m=\Ord(n^2)$~\cite{Martinez92}.)  The probability~$p_{ij}$ of an edge
appearing between any particular vertex pair $(i,j)$ is of order~$1/n$ in
such networks.  Thus, for example, in the fermionic case of the network
described by the Hamiltonian~\eref{gengenrg}, Eq.~\eref{edgeprob1} tells us
that $\e^{\Theta_{ij}}$ must be of order $n$ in a sparse graph.  The same
is also true for the bosonic network of the previous section.  This allows
us to approximate many of our expressions by ignoring terms of order 1 by
comparison with terms of order~$\e^{\Theta_{ij}}$.  We refer to such
approximations as the ``sparse limit'' or the ``classical limit,'' the
latter by analogy with the corresponding phenomenon in quantum gases at low
density.

In particular, the equivalent of Eq.~\eref{edgeprob1} for either fermionic
or bosonic graphs in the classical limit is $p_{ij} = \e^{-\Theta_{ij}}$.
For the case of Eq.~\eref{genrg}, it is
\begin{equation}
p_{ij} = \e^{-\theta_i}\e^{-\theta_j},
\label{edgeprob3}
\end{equation}
so that each edge appears with a probability that is a simple product of
``fugacities''~$\e^{-\theta_i}$ defined on each vertex.  The classical
limit of this model has been studied previously by a number of other
authors~\cite{GKK01b,CL02a,CCDM02,DMS03a,PN03}, although again developed
and justified in a different way from our presentation here; generally the
edge probability~\eref{edgeprob3} has been taken as an assumption, rather
than a derived result.

For a given distribution~$\rho(\theta)$ of~$\theta$, the expected degree of
a vertex, Eq.~\eref{distkav}, is
\begin{equation}
\av{k_i} = (n-1)\,\e^{-\theta_i} \int \e^{-\theta'} \rho(\theta')\,\d\theta',
\end{equation}
which is simply proportional to~$\e^{-\theta_i}$.  So we can produce any
desired degree distribution by choosing the corresponding distribution
for~$\theta$.

\subsection{Fixed edge counts}
\label{fixed}
Another possible choice of graph set~$\cG$ is the set of graphs with both a
fixed number of vertices~$n$ and a fixed number of edges~$m$.  Models of
this kind have been examined occasionally in the literature~\cite{BL02}
and, if we once more adopt the view of the edges in a graph as particles,
they can be considered to be the canonical ensemble of network models,
where the variable edge-count models of previous sections are the grand
canonical ensemble.  As in conventional statistical mechanics, the grand
ensemble is often simpler to work with than the canonical one, but progress
can be made sometimes be made in the canonical case by performing the sum
over all graphs regardless of edge count and introducing a Kronecker
$\delta$-symbol into the partition function to impose the edge constraint:
\begin{equation}
Z = \sum_G \delta\bigl(\widetilde{m},m(G)\bigr) \e^{-H},
\end{equation}
where $\widetilde{m}$ is the desired number of edges.  

For instance, the fixed edge-count version of the generalized random graph,
Eq.~\eref{gengenrg}, would be one in which
\begin{eqnarray}
Z &=& \sum_G \delta(\widetilde{m},m)
      \exp\biggl({-\sum_{i<j} \Theta_{ij}\sigma_{ij}}\biggr) \nonumber\\
  &=& \int_0^1\d\eta\>\e^{2\pi\ii\widetilde{m}\eta}
      \sum_{\set{\sigma_{ij}}}
      \exp\biggl({-\sum_{i<j} (\Theta_{ij}+2\pi\ii\eta) \sigma_{ij}}\biggr),
      \nonumber\\
\end{eqnarray}
where we have made use of the integral representation for the
$\delta$-function
\begin{equation}
\delta(\widetilde{m},m) =
  \int_0^1 \e^{2\pi\ii(\widetilde{m}-m)\eta}\>\d\eta.
\end{equation}
The sum over graphs is now in the form of the partition function for the
grand canonical version of the model, but with
$\Theta_{ij}\to\Theta_{ij}+2\pi\ii\eta$, giving the field parameters an
imaginary part.  Thus, from Eq.~\eref{gengenrgpart}
\begin{equation}
Z = \int_0^1 \d\eta\>\e^{2\pi\ii\widetilde{m}\eta}
    \prod_{i<j} \bigl( 1 + \e^{-(\Theta_{ij}+2\pi\ii\eta)} \bigr).
\end{equation}
In general the integral cannot be done in closed form, which is why fixed
edge-count graphs---and canonical ensembles in general---are avoided.  The
integral can in principle be carried out term by term for any finite~$n$,
but doing so is tantamount to performing the sum over all graphs with
$\widetilde{m}$ edges explicitly, so there is little to be gained by the
exercise.

It is also possible to have a bosonic graph with a fixed number of
edges---one would simply sum over the set of graphs that have
$\widetilde{m}$ edges with any number of them being permitted to fall
between any given pair of vertices.

We will not discuss further either fixed edge-counts or bosonic networks in
this paper, concentrating instead on the grand canonical fermionic ones,
which are more useful overall.  However, essentially all of the results
reported in the remainder of the paper can be generalized, with a little
work, to these other cases if necessary.

\section{More complex Hamiltonians}
\label{morecomplex}
Outside of the models described in the previous sections, and some minor
variations on them, we know of few other exponential random graph models
that are exactly solvable.  (One exception is the reciprocity model of
Holland and Leinhardt~\cite{HL81}, for which we derive an exact solution in
Sec.~\ref{reciprocity}.)  To make further progress one must turn to
approximate methods.  There are (at least) three types of techniques that
can yield approximate analytic solutions for exponential random graph
models.  The first and simplest is mean-field theory, which works well in
many cases because of the intrinsically high dimensionality of network
models; usually these models have an effective dimensionality that
increases with the number of vertices~$n$, so that the thermodynamic limit
of $n\to\infty$ also corresponds to the high dimension limit in which
mean-field theory becomes accurate.  Nonetheless, there are many
quantities, such as those depending on fluctuations, about which mean-field
theory says nothing, and for these other methods are needed.  In some cases
one can use non-perturbative approaches based on the Hubbard--Stratonovich
transform or similar integral transforms, which are very effective and
accurate but suitable only for models with Hamiltonians of specific forms
polynomial in the adjacency matrix.  More generally, one can use
perturbation theory, which may involve larger approximations (although they
are usually well controlled), but is applicable to Hamiltonians of
essentially any form.

We discuss all of these approaches here.  As an example of their
application, we use one of the oldest and best-studied of exponential
random graphs, the 2-star model.  The Hamiltonian for the 2-star model is
\begin{equation}
H = \theta m - \tsparam s,
\label{2star1}
\end{equation}
where $m$ is the number of edges in the network and $s$ is the number of
``2-stars.''  A 2-star is two edges connected to a common vertex.  (The
minus sign in front of the parameter $\tsparam$ is introduced for later
convenience.)

The quantities $m$ and $s$ can be rewritten in terms of the degree sequence
thus:
\begin{equation}
m = \half \sum_i k_i,\quad s = \half \sum_i k_i(k_i-1).
\end{equation}
Substituting these expressions into Eq.~\eref{2star1}, we can rewrite the
Hamiltonian as
\begin{equation}
H = -{J\over n-1}\sum_i k_i^2 - B\sum_i k_i,
\label{2star2}
\end{equation}
where $J=\half(n-1)\tsparam$ and $B=-\half(\theta+\tsparam)$.  (The factor
$(n-1)$ in the definition of $J$ is also introduced for convenience later
on.)

Noticing once again that $k_i=\sum_j \sigma_{ij}$, where the variables
$\sigma_{ij}$ are the elements of the adjacency matrix, we can also write
\begin{equation}
H = -{J\over n-1}\sum_{ijk} \sigma_{ij}\sigma_{ik}
    - B\sum_{ij} \sigma_{ij}.
\label{2star3}
\end{equation}
We study the 2-star model in the fermionic case in which each vertex pair
can be connected by at most a single edge, and within the grand canonical
ensemble where the total number of edges is not fixed.  Generalization to
the other cases described above is of course possible, if not always easy.

\subsection{Mean-field theory}
\label{mftsec}
The variables $\sigma_{ij}$ can be thought of as Ising spins residing on
the edges of a fully connected graph, and hence the 2-star model can be
thought of as an Ising model on the edge-dual graph of the fully connected
graph~\cite{PDFV04}.  (The edge-dual $G^*$ of a graph~$G$ is the graph in
which each edge in $G$ is replaced by a vertex in $G^*$ and two vertices in
$G^*$ are connected by an edge if the corresponding edges in $G$ share a
vertex.)  Using this equivalence, the mean-field theory of the 2-star model
can be developed in exactly the same way as for the more familiar
lattice-based Ising model.

We begin by writing out all terms in Eq.~\eref{2star3} that involve a
particular spin~$\sigma_{ij}$:
\begin{equation}
H(\sigma_{ij}) = -\sigma_{ij} \biggl[ {J\over n-1}
\sum_k (\sigma_{ik} + \sigma_{ki} + \sigma_{jk} + \sigma_{kj}) + 2B \biggl],
\end{equation}
where we have explicitly taken account of all the ways in which
$\sigma_{ij}$ can enter the first term in the Hamiltonian.  (We have also
dropped the term $2J\sigma_{ij}/(n-1)$ required to correctly count the
terms diagonal in $\sigma_{ij}$, since it vanishes in the large $n$ limit.)

Then, in classic mean-field fashion, we approximate the local field by its
average:
\begin{equation}
{J\over n-1}
  \sum_k (\sigma_{ik} + \sigma_{ki} + \sigma_{jk} + \sigma_{kj}) + 2B
  \to 4Jp + 2B,
\end{equation}
where, as before, $p=\av{\sigma}$ is the mean probability of an edge
between any pair of vertices, which is also called the \defn{connectance}
of the graph.  Then $H(\sigma_{ij}) = -(4Jp+2B)\sigma_{ij}$, and we can
write a self-consistency condition for $p$ of the form
\begin{equation}
{p\over1-p} = {\e^{-H(\sigma_{ij}=1)}\over\e^{-H(\sigma_{ij}=0)}}
  = \e^{4Jp+2B}.
\end{equation}
Rearranging, this then gives us
\begin{equation}
p = {\e^{4Jp+2B}\over1+\e^{4Jp+2B}}
  = \half\bigl[ \tanh(2Jp+B) + 1 \bigr].
\label{meanfield}
\end{equation}

\begin{figure}
\begin{center}
\resizebox{7.5cm}{!}{\includegraphics{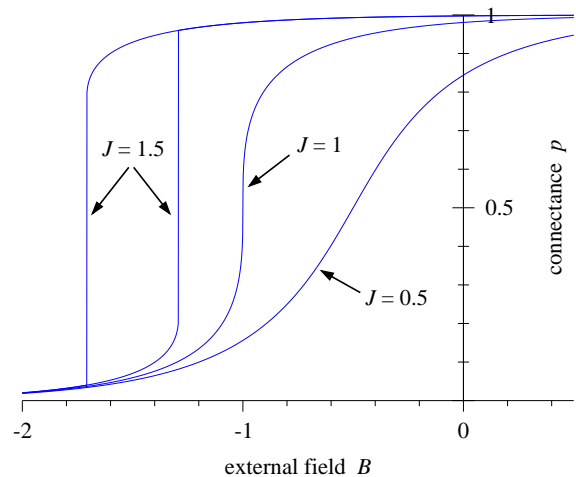}}
\end{center}
\caption{The mean-field solution for the connectance $p=\av{k}/(n-1)$
in the 2-star model from Eq.~\eref{meanfield}, for values of the coupling
$J$ below, at, and above the phase transition.  For the case $J=1.5$ we are
in the symmetry broken phase and the hysteresis loop corresponding to the
high- and low-density phases of the system is clearly visible.}
\label{mft}
\end{figure}

For $J\le1$ this equation has only one solution, but for $J>1$ there may
either be one solution or, if $B$ is sufficiently close to~$-J$, there may
be three, of which the outer two are stable.  Thus when $B$ is close to
$-J$ we have a bifurcation at $J_c=1$, a continuous phase transition to a
symmetry broken state with two phases, one of high density and one of low.
We show in Fig.~\ref{mft} a plot of the solution of~\eref{meanfield} which
displays clearly the characteristic hysteresis loop of the symmetry broken
state.

Along the ``symmetric line'' $B=-J$ there is always a solution $p=\half$ to
Eq.~\eref{meanfield} (although it may be unstable), and along this line we
can think of $p-\half$ as a standard order parameter for the model which is
zero in the high-symmetry phase and non-zero in the symmetry-broken phase.
We can define a critical exponent $\beta$ in the usual fashion by
\begin{equation}
\bigl| p - \half \bigr| \sim (J-1)^\beta,
\end{equation}
as $J\to0^+$, giving $\beta=\half$, which is the usual Ising mean-field
value and should come as no surprise, given the equivalence mentioned above
between the 2-star model and the Ising model.  One can define other
critical exponents as well, which are also found to take Ising mean-field
values.  For instance, as we showed in~\cite{PN04}, the variance~$\chi$ of
the connectance, which plays the role of a susceptibility, goes as $\chi
\sim |J-1|^{-\gamma}$ in the vicinity of the phase transition with
$\gamma=1$.

\subsection{Non-perturbative approaches}
\label{nonperturb}
We can go beyond the mean-field approximation of the previous section by
making use of techniques borrowed from many-body theory.  The developments
of this section follow closely the lines of our previous paper on this
topic~\cite{PN04}, and rather than duplicate material needlessly, the
reader is referred to that paper for details of the calculation.  Here we
merely summarize the important results.

The evaluation of the partition function for the 2-star model involves a
sum of terms of the form $\e^{k^2}$.  The study of interacting quantum
systems has taught us that such sums can be performed using the
Hubbard--Stratonovich transform.  We start by noting the well-known result
for the Gaussian integral:
\begin{equation}
\int_{-\infty}^\infty \e^{-a\phi^2}\>\d\phi = \sqrt{\pi\over a}.
\end{equation}
Making the substitutions $a\to(n-1)J$ and $\phi\to\phi_i-k_i/(n-1)$, and
rearranging, this becomes
\begin{eqnarray}
\e^{Jk_i^2/(n-1)} &=& \sqrt{(n-1)J\over\pi} \nonumber\\
   & & \hspace{-8em} \times \int_{-\infty}^\infty
       \e^{-(n-1)J\phi_i^2+2J\phi_i k_i}\,\d\phi.
\end{eqnarray}
Then the partition function is
\begin{eqnarray}
Z &=& \sum_G \exp\biggl( -{J\over n-1}\sum_i k_i^2 - B\sum_i k_i \biggr)
      \nonumber\\
  &=& \biggl[{(n-1)J\over\pi}\biggr]^{n/2}
    \int\cD\vphi\>\exp\Bigl(-(n-1)J\sum_i\phi_i^2\Bigr)\nonumber\\
  & & \hspace{3em} \times \sum_G \exp\Bigl(\sum_i (2J\phi_i+B)k_i\Bigr),
\label{2starpart1}
\end{eqnarray}
where we have interchanged the order of sum and integral.

The sum over graphs now has precisely the form of the partition function
sum for the model of Eq.~\eref{genrg}, and from Eq.~\eref{genrgpart} we can
thus immediately write down the partition function
\begin{equation}
Z = \int\cD\vphi\>\e^{-\cH(\vphi)},
\label{ft}
\end{equation}
where the quantity
\begin{eqnarray}
\cH(\vphi) &=& (n-1)J\sum_i \phi_i^2 - \half\sum_{i\ne j}
                  \ln \bigl( 1 + \e^{2J(\phi_i+\phi_j)+2B}\bigr) \nonumber\\
           & & {} - \half n\ln\bigl((n-1)J\bigr),
\label{effective}
\end{eqnarray}
is called the effective Hamiltonian.

Thus we have completed the partition function sum for the 2-star model, but
at the expense of introducing the auxiliary fields $\set{\phi_i}$ which
must be integrated out to complete the calculation.  The integral cannot,
as far as we are aware, be evaluated exactly in closed form but, as we
showed in~\cite{PN04}, it can be evaluated approximately using a
saddle-point expansion, with the result that the free energy of the 2-star
model is given to leading order in the expansion by
\begin{eqnarray}
F &=& n(n-1)J\phi_0^2 - \half n(n-1) \ln\bigl( 1 + \e^{4J\phi_0+2B} \bigr)
        \nonumber\\
  & & \quad {} + \half(n-1)\ln\bigl(1-2J\phi_0(1-\phi_0)\bigr),
\end{eqnarray}
where
\begin{equation}
\phi_0 = \half\bigl[ \tanh\bigl(2J\phi_0+B\bigr) + 1 \bigr]
\label{meanfield2}
\end{equation}
is the position of the saddle point, i.e.,~the maximum of the Hamiltonian
on the real-$\phi$ line.

Note that Eq.~\eref{meanfield2} is identical to the mean-field equation,
Eq.~\eref{meanfield}, for the connectance~$p$ of the 2-star model.  Thus,
$\phi_0$~is the connectance of the model within the mean-field
approximation and the saddle-point expansion, as is typically the case in
such calculations, is an expansion about the mean-field solution.

From the free energy we can derive a number of quantities of interest.  We
showed in~\cite{PN04}, for instance, that the variance of vertex degree in
the model is given by
\begin{equation}
\langle k^2 \rangle - \av{k}^2 =
  (n-1)\frac{\phi_0(1-\phi_0)}{1-2J\phi_0(1-\phi_0)},
\label{variance}
\end{equation}
which has a gradient discontinuity but no divergence at the phase
transition.  (This quantity is, by contrast, zero everywhere within
mean-field theory.)

\subsection{Perturbation theory}
\label{perturbation}
Exponential random graphs also lend themselves naturally to treatment using
perturbation theory.  Here we describe the simplest such theory, which is
roughly equivalent to the high-temperature expansions of conventional
thermal statistical mechanics.  Expansions of this type have been examined
previously by Burda~\etal~\cite{BJK04a,BJK04b} for Strauss's model of a
transitive network~\cite{FS86,Strauss86}.  Here we develop the theory
further for general exponential random graphs.

The fundamental idea of perturbation theory for random graphs is the same
as for other perturbative methods: we expand about a solvable model in
powers of the coupling parameters $\theta_i$ in the Hamiltonian.  We write
the Hamiltonian for the full model in the form $H = H_0 + H_1$, where $H_0$
is the Hamiltonian for the solvable model and $H_1$ takes whatever form is
necessary to give the correct expression for~$H$.  Then the partition
function is~\cite{BJK04a,BJK04b}
\begin{equation}
Z = \sum_G \e^{-(H_0+H_1)}
  = Z_0 \sum_G {\e^{-H_0}\over Z_0}\,\e^{-H_1}
  = Z_0 \av{\e^{-H_1}}_0,
\label{perturb}
\end{equation}
where $Z_0=\sum_G \e^{-H_0}$ is the partition function for the unperturbed
Hamiltonian, and $\av{\dots}_0$ indicates an ensemble average in the
unperturbed model.

The only case that has been investigated in any detail is the one where we
expand around a random graph, $H_0 = \theta m$, so that the averages in
Eq.~\eref{perturb} are averages in the ensemble of the random graph.  (It
is possible for $\theta$ to be zero, so this choice for $H_0$ does not
place any restriction on the form of the overall Hamiltonian.  If
$\theta=0$ then the expansion is precisely equivalent to an ordinary
high-temperature series.)  However, for Hamiltonians~$H$ that give
significant probability to networks substantially different from random
graphs, the perturbation theory cannot be expected to give accurate answers
at low order.  In theory there is no reason why one could not expand about
some other solvable case, although no such calculations have been done as
far as we are aware.  One obvious possibility, which we do not pursue here,
is to expand around one of the generalized random graph forms,
Eqs.~\eref{genrg} and~\eref{gengenrg}.

Typically, to make progress with Eq.~\eref{perturb}, we will expand the
exponential in a power series of the form
\begin{equation}
{Z\over Z_0} = \sum_{k=0}^\infty {(-1)^k\over k!} \av{H_1^k}_0.
\label{series}
\end{equation}
In practice, $H_1$~normally contains a coupling constant, such as the
constant~$J$ in the 2-star model of Eq.~\eref{2star2}, and thus our
expression for the perturbed partition function is an expansion in powers
of the coupling.

In this section, we apply the perturbation method to two example models.
First, we study a simple model proposed about a quarter of a century ago by
Holland and Leinhardt~\cite{HL81}, which is exactly solvable by this
method.  Then we illustrate the application of the method to the 2-star
model and compare its performance against the approximate saddle-point
expansion results of the previous section.

\subsubsection{The reciprocity model}
\label{reciprocity}
Our first example of perturbation theory is a directed graph model.  In the
real world, many directed graphs display the phenomenon of reciprocity: a
directed edge running from vertex~A to vertex~B predisposes the network to
have an edge running from B to~A as well.  Put another way, the network has
a higher fraction of vertex pairs that are joined in both directions than
one would expect on the basis of chance (``mutual dyads'' in the parlance
of social network analysis).  Behavior of this kind is seen, for example,
in the world wide web, email networks, and neural and metabolic
networks~\cite{EM02,NFB02,GL04}.

Holland and Leinhardt~\cite{HL81} have proposed a exponential random graph
model of reciprocity, which we study here in a simplified version.  As we
now show, the perturbation expansion for this model can be written down to
all orders and resummed to give an exact expression for the partition
function.

The Hamiltonian for the model is
\begin{equation}
H = H_0 + H_1 = \theta m - \hlparam r,
\label{hamrecip}
\end{equation}
where $m$ is the total number of (directed) edges in the graph, and $r$ is
the number of vertex pairs with edges running between them in both
directions.  The unperturbed Hamiltonian~$H_0$ is that of an undirected
random graph (Sec.~\ref{directedgraphs}) with partition function given by
Eq.~\eref{directedz} and each directed edge present with independent
probability $p=(\e^\theta+1)^{-1}$.  The perturbation~$H_1$ can be written
in terms of the adjacency matrix~\eref{adjacencydirected} as
\begin{equation}
H_1 = -\hlparam r = -\hlparam\sum_{i<j} \sigma_{ij}\sigma_{ji}.
\end{equation}
Then the perturbation series, Eq.~\eref{series}, for the full Hamiltonian
is
\begin{equation}
{Z\over Z_0} = \sum_{k=0}^\infty {\hlparam^k \over k!} \av{r^k}_0,
\label{recippartition}
\end{equation}
with
\begin{equation}
\av{r^k}_0 = \sum_{i_1<j_1}\ldots\sum_{i_k<j_k}
\av{\sigma_{i_1j_1}\sigma_{j_1i_1}\ldots\sigma_{i_kj_k}\sigma_{j_ki_k}}_0.
\label{correlator}
\end{equation}
Thus the partition function is written as an expansion in powers
of~$\hlparam$ whose coefficients are correlation functions of elements of
the adjacency matrix, calculated within the ordinary random graph.  If we
can evaluate these correlation functions, at least up to some finite order,
we can also evaluate the perturbed partition function.

Since all edges are present or absent independently of one another in the
random graph, the correlation functions factor:
\begin{equation}
\av{\sigma_{12}\sigma_{21}\sigma_{34}\sigma_{43}}_0
  = \av{\sigma_{12}}_0 \av{\sigma_{21}}_0
    \av{\sigma_{34}}_0 \av{\sigma_{43}}_0 
  = p^4,
\end{equation}
and so forth.  The only exception is in cases where two or more of the
elements $\sigma_{ij}$ being averaged are the same.  In that case, noting
that $\sigma_{ij}^n = \sigma_{ij}$ for any~$n$, we have results like
\begin{equation}
\av{\sigma_{12}\sigma_{21}\sigma_{12}\sigma_{21}}_0
  = \av{\sigma_{12}}_0 \av{\sigma_{21}}_0
  = p^2.
\end{equation}
To evaluate expressions such as~\eref{correlator}, therefore, we need to
count the number of independent elements $\sigma_{ij}$ that appear in each
term.  This can be difficult for some models, but for the reciprocity model
it is quite straightforward.  The question we need to answer is this: if we
choose $k$ pairs of vertices $(i,j)$ from the ${n\choose2}$ possible pairs,
with duplication allowed, how many ways are there of choosing them such
that exactly $q$ pairs will be distinct?  Each such way makes a
contribution of $p^{2q}$ to the partition function~\eref{recippartition}.

Let $a_{k,q}$ be the number of ways of choosing the pairs such at a
\emph{particular} set of $q$ distinct pairs are chosen at least once each.
Note that $a_{k,1}=1$ for all~$k$.  Then, from Eqs.~\eref{recippartition}
and~\eref{correlator}
\begin{eqnarray}
{Z\over Z_0} &=& 1 + \sum_{k=1}^\infty {\hlparam^k\over k!}
                 \sum_{q=1}^{n\choose2} {{n\choose2}\choose q}
                 a_{k,q} p^{2q} \nonumber\\
             &=& 1 + \sum_{q=1}^{n\choose2} {{n\choose2}\choose q}
                 g_q(\hlparam) p^{2q},
\label{comb}
\end{eqnarray}
where the function
\begin{equation}
g_q(z) = \sum_{k=1}^\infty {z^k\over k!}\,a_{k,q}
\end{equation}
is the exponential generating function for~$a_{k,q}$.

Now the number of ways of choosing $k$ pairs such that all choices are made
from a particular set of size~$q$, but without the constraint that each
pair in the set appear at least once, is just $q^k$.  Thus $\sum_{m=1}^q
{q\choose m} a_{k,m} = q^k$, or equivalently
\begin{equation}
a_{k,q} = q^k - \sum_{m=1}^{q-1} {q\choose m} a_{k,m}.
\end{equation}
Multiplying by $z^k/k!$ and summing over $k=1\ldots\infty$, this gives
\begin{equation}
g_q(z) = \e^{qz} - 1 - \sum_{m=1}^{q-1} {q\choose m} g_m(z),
\label{induction}
\end{equation}
which immediately implies that $g_q(z) = (e^z-1)^q$, by induction
on~\eref{induction} with the initial condition $g_1(z) = \sum_{k=1}^\infty
z^k a_{k,1}/k! = \e^z-1$.

Substituting this result into Eq.~\eref{comb} then gives us our solution:
\begin{equation}
{Z \over Z_0} = \sum_{q=1}^{n\choose2} {{n\choose2}\choose q}
                (\e^\hlparam-1)^q p^{2q}
              = \bigl[1+(\e^\hlparam-1)p^2\bigr]^{n\choose 2}.
\end{equation}
Or, making use of Eqs.~\eref{bernoulli} and~\eref{defsp}
\begin{equation}
Z = \biggl[ {1+(\e^\hlparam-1)p^2\over1-p} \biggr]^{n\choose2},
\label{hlz}
\end{equation}
and $F=-\ln Z$ in the normal fashion.

From these expressions we can, for instance, obtain the mean number of
edges~$\av{m}$ and the mean number~$\av{r}$ of pairs of vertices connected
by edges running both ways from
\begin{equation}
\av{m} = {\partial F\over\partial\theta}
       = p(p-1) {\partial F\over\partial p},\quad
\av{r} = -{\partial F\over\partial\hlparam}.
\end{equation}
A quantity of interest in directed networks is the
\defn{reciprocity}~\cite{NFB02}, which is the fraction of edges that are
reciprocated.  This quantity is found to be on the order of tens of percent
in networks such as the world wide web.  The reciprocity for the model of
Holland and Leinhardt~is
\begin{equation}
{2\av{r}\over\av{m}} = {p\e^\hlparam\over1-p+p\e^\hlparam}.
\end{equation}

\begin{figure}
\resizebox{\figurewidth}{!}{\includegraphics{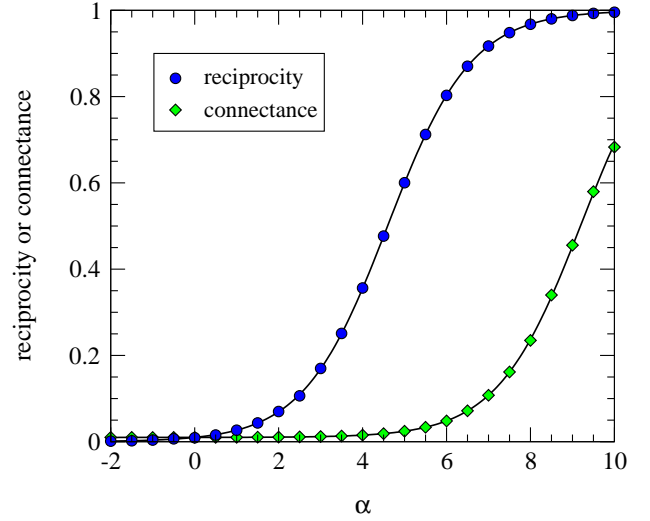}}
\caption{The reciprocity and connectance of the model of Holland and
Leinhardt~\cite{HL81} for $p=0.01$.  The solid lines represent the exact
solution, Eq.~\eref{hlz}, and the points are Monte Carlo simulation results
for systems of $n=1000$ vertices.}
\label{rp}
\end{figure}

In Fig.~\ref{rp}, we show the reciprocity, along with the connectance of
the network, as a function of $\hlparam$ for the case $p=0.01$.  There is
no phase transition or other unexpected behavior in this model: the
measured properties are smooth functions of the independent parameters.
Notice that there is a substantial range of values of $\hlparam$ over which
the connectance is low and the graph realistically sparse, but the
reciprocity is still high, with values similar to those seen in real
networks.

\subsubsection{Example 2: The 2-star model}
As our second example of the application of perturbation theory, we return
to the 2-star model introduced at the beginning of Sec.~\ref{morecomplex}.
Unlike the case of the reciprocity model in the preceding section,
perturbation theory does not lead to an exact solution of the 2-star model
but, as we now show, we can get an approximate solution by studying the
perturbation expansion to finite order---a different approximation from the
saddle-point expansion of Sec.~\ref{nonperturb}.

We divide the Hamiltonian $H=\theta m - \tsparam s$ into an unperturbed
part $H_0=\theta m$, which is the normal Bernoulli random graph, and a
perturbation Hamiltonian $H_1=-\tsparam s$.  Then, following
Eq.~\eref{series}, the partition function for the full model is given by
\begin{equation}
{Z\over Z_0} = \sum_{l=0}^\infty {\tsparam^l\over l!} \av{s^l}_0.
\label{2starpartition}
\end{equation}
The number of 2-stars is
\begin{equation}
s = \sum_i \sum_{j<k} \sigma_{ij}\sigma_{ik},
\label{n2stars}
\end{equation}
and therefore
\begin{equation}
\av{s^l}_0 = \sum_{i_1,j_1<k_1}\!\ldots\!\sum_{i_l,j_l<k_l}
             \av{\sigma_{i_1j_1}\sigma_{i_1k_1}\ldots
             \sigma_{i_lj_l}\sigma_{i_lk_l}}_0.
\label{2starcorrelator}
\end{equation}

Our strategy is to evaluate the series~\eref{2starpartition} up to some
finite order in $\tsparam$ to get an approximate solution for $Z$, but
there is a problem.  Each term in the series corresponds to states of the
graph that have the corresponding number of 2-stars: the term in
$\av{s}_0$, for instance, counts the number of graphs that have a 2-star in
any position in the graph.  This is not enough for our purposes however.
Realistic graphs will have not a finite number but a finite \emph{density}
of 2-stars in them, and the number of such graphs is counted by terms that
appear at infinite order in the perturbation expansion in the limit
$n\to\infty$.  So, without going to infinite order as we did in the
reciprocity model, we are never going to get meaningful results from our
expansion.

Similar problems appear in ordinary statistical mechanics and the solution
is well known.  Instead of expanding the partition function, we form an
expansion for the free energy.  We can write the free energy as
\begin{equation}
F = -\ln Z = -\ln Z_0 - \ln {Z\over Z_0} = F_0 + F_1,
\end{equation}
where $F_0$ is the free energy of the unperturbed network and
$F_1=-\ln(Z/Z_0)$.  Now we expand $F_1$ as a power series in~$\tsparam$ of
the form
\begin{equation}
F_1 = - \tsparam f_1 - {\tsparam^2\over2!} f_2
      - {\tsparam^3\over3!} f_3 - \ldots,
\end{equation}
where we have made use of the fact that $F_1=0$ when $\tsparam=0$.
Substituting into $Z/Z_0=\e^{-F_1}$, we get
\begin{equation}
{Z\over Z_0} = 1 + \tsparam f_1 + {\tsparam^2\over2!} (f_2 + f_1^2)
               + {\tsparam^3\over3!} (f_3 + 3 f_2 f_1 + f_1^3)
               + \Ord(\tsparam^4),
\end{equation}
and comparing terms with Eq.~\eref{2starpartition}, we find
\begin{subequations}
\label{cumulants}
\begin{eqnarray}
f_1 &=& \av{s}_0,\\
\label{cumulantsb}
f_2 &=& \av{s^2}_0 - f_1^2,\\
f_3 &=& \av{s^3}_0 - 3f_2f_1 - f_1^3,
\end{eqnarray}
\end{subequations}
and so forth.  These are the \defn{cumulants} of $s$ within the ensemble
defined by the unperturbed network.  If we expand $s$ in the form of
Eq.~\eref{n2stars} then they are connected correlations of elements of the
adjacency matrix---``connected'' because individual elements of the
adjacency matrix are uncorrelated, so that all terms in the cumulants
vanish unless they involve sets of 2-stars that share one or more edges.
(Note that sharing a vertex, as in the more familiar spin models of
traditional statistical mechanics, is not a sufficient condition for being
connected.  The fundamental degrees of freedom in a network are the edges.)

We will proceed then as follows.  We calculate the free energy $F_1$ in
terms of connected correlations up to some finite order in~$\tsparam$ and
from this we calculate the partition function $Z = Z_0\e^{-F_1}$.  Even
though $F_1$ is known only to finite order, our expression for $Z$ will
include terms with all powers of the connected correlations in it, via the
expansion of the exponential, and hence will include graphs with not only a
finite number but a finite density of 2-stars.  This idea, which will be
routine for those familiar with conventional diagrammatic many-body theory,
is entirely general and can be applied to any model, not just the 2-star
model.  In essence, the series given by $\e^{-F_1}$ is a partial
resummation to all orders of the partition function, including some but not
all of the contributions to $Z$ from disconnected correlations of
arbitrarily high order.

\begin{figure}
\resizebox{\figurewidth}{!}{\includegraphics{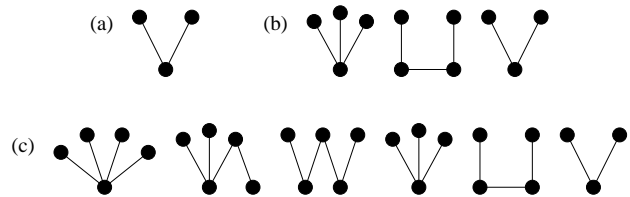}}
\caption{The diagrams contributing to the first three orders in the
perturbation expansion of the free energy of the 2-star model in powers
of~$\tsparam$.}
\label{condia}
\end{figure}

Let us see how the calculation proceeds for the case of the 2-star model,
to order~$\tsparam^3$, as above.  The leading $\Ord(\tsparam)$ term in
$F_1$ is simple:
\begin{equation}
f_1 = \av{s}_0 = \sum_i \sum_{j<k} \av{\sigma_{ij}\sigma_{ik}}_0
    = n{n-1\choose2} p^2.
\label{f1}
\end{equation}
Since we are primarily interested in large networks, we can approximate
this expression by its value to leading order in $n$, which is $\half
n^3p^2$.

The second term, at order~$\tsparam^2$, is more complicated because there
are several different ways in which two 2-stars may combine to share one or
more edges.  In order to keep track of these different contributions, we
make use of a diagrammatic representation similar to that employed by
Burda~\etal\ for Strauss's transitivity model~\cite{BJK04a}.
Figure~\ref{condia}a shows the single diagram contributing to~$f_1$, which
gives the result in Eq.~\eref{f1}.  Figure~\ref{condia}b shows the three
diagrams that contribute to~$f_2$.  It is an assumption of our notation
that each edge that appears in a diagram is distinct.  Thus the third
diagram in Fig.~\ref{condia}b, which represents the case in which the two
2-stars fall on top of one another, must be depicted separately, rather
than being considered a special case of the first diagram.  This turns out
to be a good idea, since this term has a different functional form from the
first diagram, and neither diagram is necessarily negligible by comparison
with the other.

In general the basic ``Feynman rules'' for interpreting the diagrams are:
\begin{enumerate}
\setlength{\itemsep}{0pt}
\setlength{\topsep}{0pt}
\item each edge contributes a factor of~$p$;
\item each vertex contributes a factor of~$n$;
\item the numerical multiplier is the number of distinct ways in which the
diagram can be decomposed into overlapping 2-stars such that each edge
occurs at least once, divided by the symmetry factor for the diagram.  (The
symmetry factor is the number of distinct permutations of the vertices that
leave the diagram unchanged.)
\end{enumerate}
Then for the connected correlation functions one must subtract all other
ways of composing lower order diagrams to make the given diagram, as in
Eq.~\eref{cumulants}.

To see how these rules work in practice, let us apply them to the first
diagram in Fig.~\ref{condia}b.  This diagram has four vertices and three
edges, which gives a factor of $n^4p^3$, by the first two rules.  The
diagram can be decomposed into two 2-stars in 6 different ways, but the
symmetry factor is also 6, so we end up with $n^4p^3\times6/6=n^4p^3$.  The
contribution to the diagram from the term $-f_1^2$ in Eq.~\eref{cumulantsb}
is $-n^4p^4$, so the final value of the diagram is $n^4(p^3-p^4)$ to
leading order in~$n$.  Proceeding in a similar fashion, the other diagrams
of Fig.~\ref{condia}b contribute $n^4(p^3-p^4)$ and $\half n^3(p^2-p^4)$,
respectively.  The diagrams for the $\Ord(\tsparam^3)$ term are shown in
Fig.~\ref{condia}c, and are more complicated, but routine to evaluate using
the rules above.  The final expressions for the $f$s are:
\begin{subequations}
\begin{eqnarray}
f_1 &=& \half n^3 p^2,\\
f_2 &=& \half n^3 (1-p)p^2 (1 + 4np),\\
f_3 &=& \half n^3 (1-p)p^2 (1 + 14np + 29n^2p^2 - 43n^2p^3).\nonumber\\
\end{eqnarray}
\end{subequations}
Note that we have retained the leading order terms in $n$ separately at
each order in~$p$, since we have no knowledge \textit{a priori} about the
relative magnitude of $n$ and~$p$.  In a sparse graph, we expect that $p$
will be of order $1/n$, in which case it may be possible to neglect some
terms.

Once we have the expansion of~$F_1$, it is straightforward to calculate
statistical averages from derivatives of the free energy in the normal
fashion.  For example, the expected number of 2-stars in the network is
given by
\begin{equation}
\av{s} = -{\partial F\over\partial\tsparam}
       = -{\partial F_1\over\partial\tsparam}
       = f_1 + \tsparam f_2 + \half \tsparam^2 f_3 + \Ord(\tsparam^3).
\end{equation}
And the expected number of edges is
\begin{eqnarray}
\av{m} &=& {\partial F\over\partial\theta}
        =  p(p-1) \biggl( {\partial F_0\over\partial p}
           + {\partial F_1\over\partial p} \biggr) \nonumber\\
	&=& \half n^2 p + n^3(1-p)p^2\tsparam \bigl[
            1 + \half(1+6np-8np^2)\tsparam\nonumber\\
	& & \hspace{-1em} {} + \mbox{$\frac16$}
        (1+21np+58n^2p^2-180n^2p^3+129n^2p^4)\tsparam^2\bigr].\nonumber\\
\end{eqnarray}

\begin{figure}
\resizebox{\figurewidth}{!}{\includegraphics{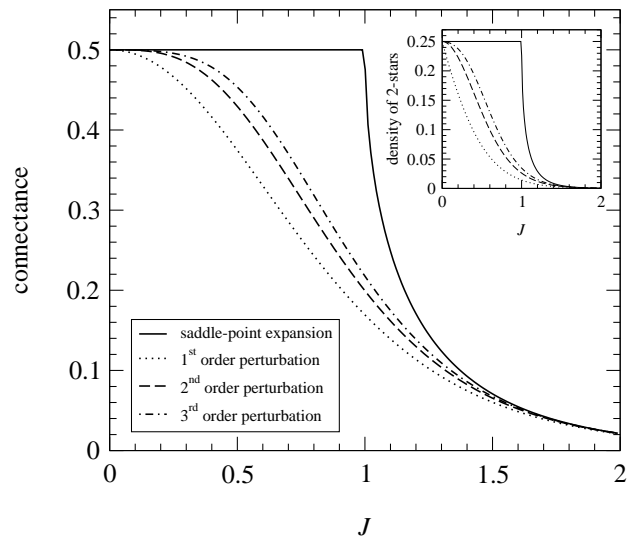}}
\caption{The connectance of the 2-star model calculated from the
saddle-point expansion of Sec.~\ref{nonperturb} (solid line), and from the
first- (dotted line), second- (dashed line), and third-order (dot-dashed
line) perturbation expansions.  The calculations were performed along the
symmetric line $B=-J$ of Sec.~\ref{mftsec}, where the half-filled graph
with connectance~$\half$ is always a solution of the mean-field
equation~\eref{meanfield2}.  For $J>1$ there exist two symmetry-equivalent
stable solutions in addition to the half-filled graph.  We show only the
sparser of the two.  Inset: the density of 2-stars in the same model.}
\label{2star}
\end{figure}

In Fig.~\ref{2star}, we show the connectance $2\av{m}/n^2$ and the density
of 2-stars $2\av{s}/n^3$ calculated from the saddle-point method of
Sec.~\ref{nonperturb} and from the expressions above, at first, second, and
third order.  As the figure shows, the perturbation expansion agrees with
the non-perturbative method at high and low values
of~$J=\half(n-1)\tsparam$, and markedly better for the third-order
approximation than for the first- and second-order ones.  However, in the
region of the phase transition at $J_c=1$ the agreement is poor, as we
would expect.  In this region there will be large critical fluctuations and
hence contributions to the free energy from large connected diagrams that
are entirely missing from our series expansion.  Presumably by extending
the perturbation series we can derive successively more accurate answers in
the critical region.  We also note that the perturbation expansion gives
results only for the sparse phase in the symmetry-broken region.

We have here studied in detail two examples of the treatment of exponential
random graphs by perturbation theory (and another can be found in
Ref.~\cite{BJK04a}).  The techniques we have used, however, are entirely
general and diagrammatic theories similar to these, with similarly simple
``Feynman rules,'' can be derived for other examples as well.

\section{Conclusions}
In this paper we have discussed exponential random graphs, which in both a
figurative and a quantitative sense play the role of a Boltzmann ensemble
for the study of networks.  Exponential random graphs are a formally
well-founded framework for making predictions about the expected properties
of networks given specific measurements of properties of those networks.
We have shown in this paper how they can be derived in moderately rigorous
fashion from maximum entropy assumptions about probability distributions
over graph ensembles.

We have given many examples of particular calculations using exponential
random graphs, starting with simple random graph models that have linear
Hamiltonians, many of which have been presented previously by other
authors, albeit it with rather different motivation.  In most cases these
linear models can be solved exactly, meaning that we can derive the
partition function or equivalently the free energy of the graph ensemble
exactly in the limit of large system size.

For nonlinear Hamiltonians it appears possible to find exact solutions only
rarely, but we have been able to find approximation solutions in several
cases using a number of different methods.  Taking the particular example
of the 2-star model, we have shown how its behavior can be understood using
mean-field theory, perturbation theory, and non-perturbative methods based
on the Hubbard--Stratonovich transform.  We have also given one example,
the reciprocity model of Holland and Leinhardt, that is exactly solvable by
evaluating its perturbation expansion to all orders.

The results presented in this paper are only a tiny fraction of what can be
done with exponential random graphs.  There are many interesting
challenges, both practical and mathematical, posed by this class of models.
Exploration of the behavior and predictions of specific models as functions
of their free parameters, development of other approximate solution
methods, or expansion of those presented here, and the development of
models to study network phenomena of particular interest, such as
vertex--vertex correlations, effects of hidden variables, effects of
degree distributions, and transitivity, are all excellent directions for
further research.  We hope to see some of these topics pursued in the near
future.

\begin{acknowledgments}
The authors thank Julian Besag, Mark Handcock, and Pip Pattison for useful
conversations.  This work was supported in part by the National Science
Foundation under grant number DMS--0234188 and by the James S. McDonnell
Foundation.
\end{acknowledgments}


\begin{thebibliography}{10}
\expandafter\ifx\csname url\endcsname\relax
  \def\url#1{\texttt{#1}}\fi
\expandafter\ifx\csname urlprefix\endcsname\relax\def\urlprefix{URL }\fi

\bibitem{Strogatz01}
S.~H. Strogatz, Exploring complex networks. \textit{Nature} \textbf{410},
  268--276 (2001).

\bibitem{AB02}
R.~Albert and A.-L. Barab\'asi, Statistical mechanics of complex networks.
  \textit{Rev. Mod. Phys.} \textbf{74}, 47--97 (2002).

\bibitem{DM02}
S.~N. Dorogovtsev and J.~F.~F. Mendes, Evolution of networks. \textit{Advances
  in Physics} \textbf{51}, 1079--1187 (2002).

\bibitem{Newman03d}
M.~E.~J. Newman, The structure and function of complex networks. \textit{SIAM
  Review} \textbf{45}, 167--256 (2003).

\bibitem{WS98}
D.~J. Watts and S.~H. Strogatz, Collective dynamics of `small-world' networks.
  \textit{Nature} \textbf{393}, 440--442 (1998).

\bibitem{BA99b}
A.-L. Barab\'asi and R.~Albert, Emergence of scaling in random networks.
  \textit{Science} \textbf{286}, 509--512 (1999).

\bibitem{KRL00}
P.~L. Krapivsky, S.~Redner, and F.~Leyvraz, Connectivity of growing random
  networks. \textit{Phys. Rev. Lett.} \textbf{85}, 4629--4632 (2000).

\bibitem{DMS00}
S.~N. Dorogovtsev, J.~F.~F. Mendes, and A.~N. Samukhin, Structure of growing
  networks with preferential linking. \textit{Phys. Rev. Lett.} \textbf{85},
  4633--4636 (2000).

\bibitem{HL81}
P.~W. Holland and S.~Leinhardt, An exponential family of probability
  distributions for directed graphs. \textit{J. Amer. Stat. Assoc.}
  \textbf{76}, 33--50 (1981).

\bibitem{Besag74}
J.~E. Besag, Spatial interaction and the statistical analysis of lattice
  systems. \textit{J. Roy. Stat. Soc. B} \textbf{36}, 192--236 (1974).

\bibitem{Frank81}
O.~Frank, An exponential family of probability distributions for directed
  graphs: Comment. \textit{J. Amer. Stat. Assoc.} \textbf{76}, 58--59 (1981).

\bibitem{FS86}
O.~Frank and D.~Strauss, Markov graphs. \textit{J. Amer. Stat. Assoc.}
  \textbf{81}, 832--842 (1986).

\bibitem{Strauss86}
D.~Strauss, On a general class of models for interaction. \textit{SIAM Review}
  \textbf{28}, 513--527 (1986).

\bibitem{WP96}
S.~Wasserman and P.~Pattison, Logit models and logistic regressions for social
  networks: {I.} {A}n introduction to {M}arkov random graphs and p$^*$.
  \textit{Psychometrika} \textbf{61}, 401--426 (1996).

\bibitem{AWC99}
C.~Anderson, S.~Wasserman, and B.~Crouch, A p* primer: Logit models for social
  networks. \textit{Social Networks} \textbf{21}, 37--66 (1999).

\bibitem{BL02}
J.~Berg and M.~L{\"a}ssig, Correlated random networks. \textit{Phys. Rev.
  Lett.} \textbf{89}, 228701 (2002).

\bibitem{PN03}
J.~Park and M.~E.~J. Newman, The origin of degree correlations in the
  {I}nternet and other networks. \textit{Phys. Rev. E} \textbf{68}, 026112
  (2003).

\bibitem{PN04}
J.~Park and M.~E.~J. Newman, Solution of the 2-star model of a network.
  Preprint cond-mat/0405457 (2004).

\bibitem{BJK04a}
Z.~Burda, J.~Jurkiewicz, and A.~Krzywicki, Network transitivity and matrix
  models. \textit{Phys. Rev. E} \textbf{69}, 026106 (2004).

\bibitem{BJK04b}
Z.~Burda, J.~Jurkiewicz, and A.~Krzywicki, Perturbing general uncorrelated
  networks. Preprint cond-mat/0401310 (2004).

\bibitem{PDFV04}
G.~Palla, I.~Der\'enyi, I.~Farkas, and T.~Vicsek, Statistical mechanics of
  topological phase transitions in networks. \textit{Phys. Rev. E} \textbf{69},
  046117 (2004).

\bibitem{Snijders02}
T.~A.~B. Snijders, {M}arkov chain {M}onte {C}arlo estimation of exponential
  random graph models. \textit{Journal of Social Structure} \textbf{2}(2)
  (2002).

\bibitem{SR51}
R.~Solomonoff and A.~Rapoport, Connectivity of random nets. \textit{Bulletin of
  Mathematical Biophysics} \textbf{13}, 107--117 (1951).

\bibitem{ER59}
P.~Erd\H{o}s and A.~R\'enyi, On random graphs. \textit{Publicationes
  Mathematicae} \textbf{6}, 290--297 (1959).

\bibitem{ER60}
P.~Erd\H{o}s and A.~R\'enyi, On the evolution of random graphs.
  \textit{Publications of the Mathematical Institute of the Hungarian Academy
  of Sciences} \textbf{5}, 17--61 (1960).

\bibitem{FFF99}
M.~Faloutsos, P.~Faloutsos, and C.~Faloutsos, On power-law relationships of the
  internet topology. \textit{Computer Communications Review} \textbf{29},
  251--262 (1999).

\bibitem{Kleinberg99b}
J.~M. Kleinberg, S.~R. Kumar, P.~Raghavan, S.~Rajagopalan, and A.~Tomkins, The
  {W}eb as a graph: Measurements, models and methods. In \textit{Proceedings of
  the International Conference on Combinatorics and Computing}, number 1627 in
  Lecture Notes in Computer Science, pp. 1--18, Springer, Berlin (1999).

\bibitem{Martinez92}
N.~D. Martinez, Constant connectance in community food webs. \textit{American
  Naturalist} \textbf{139}, 1208--1218 (1992).

\bibitem{GKK01b}
K.-I. Goh, B.~Kahng, and D.~Kim, Universal behavior of load distribution in
  scale-free networks. \textit{Phys. Rev. Lett.} \textbf{87}, 278701 (2001).

\bibitem{CL02a}
F.~Chung and L.~Lu, Connected components in random graphs with given degree
  sequences. \textit{Annals of Combinatorics} \textbf{6}, 125--145 (2002).

\bibitem{CCDM02}
G.~Caldarelli, A.~Capocci, P.~De~Los~Rios, and M.~A. Mu{\~n}oz, Scale-free
  networks from varying vertex intrinsic fitness. \textit{Phys. Rev. Lett.}
  \textbf{89}, 258702 (2002).

\bibitem{DMS03a}
S.~N. Dorogovtsev, J.~F.~F. Mendes, and A.~N. Samukhin, Principles of
  statistical mechanics of random networks. \textit{Nucl. Phys. B}
  \textbf{666}, 396--416 (2003).

\bibitem{EM02}
J.-P. Eckmann and E.~Moses, Curvature of co-links uncovers hidden thematic
  layers in the world wide web. \textit{Proc. Natl. Acad. Sci. USA}
  \textbf{99}, 5825--5829 (2002).

\bibitem{NFB02}
M.~E.~J. Newman, S.~Forrest, and J.~Balthrop, Email networks and the spread of
  computer viruses. \textit{Phys. Rev. E} \textbf{66}, 035101 (2002).

\bibitem{GL04}
D.~Garlaschelli and M.~I. Loffredo, Patterns of link reciprocity in directed
  networks. Preprint cond-mat/0404521 (2004).

\end{thebibliography}
\end{document}